# Comment on "Loophole-free Bell inequality violation using electron spins separated by 1.3 kilometers".

Louis Vervoort, 05.02.2016

*Institut National de Recherche Scientifique, and Minkowski Institute, Montreal, Canada*

**Abstract**. In a recent article Hensen et al. [Nature 526, 682 (2015)] report on a sophisticated Bell experiment, simultaneously closing, for the first time, loopholes for local hidden-variable theories (HVTs). The authors claim that 'local realism' has been refuted, under certain natural assumptions. The aim of the present Comment is twofold. First, it is urged that the class of local HVTs that is eliminated by the reported experiment should be specified in greater detail than is done by the authors. Second, it is argued that the class of local HVTs that still survives is wide and natural. For instance, hidden variables describing a 'background' field can exploit the freedom-of-choice loophole, which cannot be closed for this type of hidden variables. The dynamics of such background-based theories can be illustrated by existing systems, e.g. from fluid mechanics.

In a recent article [1], a team led by Ronald Hanson reports on a remarkable Bell experiment, for the first time simultaneously closing the experimental loopholes for local hidden-variable theories (HVTs). A short time later a second [2] and third [3] article have been published by other teams, drawing largely the same conclusions. The authors of Ref. [1] claim a very general result – essentially the refutation of any 'local-realist' theory (under some specifications, cf. Claim 1 below). The aim of the present Comment is twofold. First, it is urged that the class of local HVTs that is eliminated by [1] should be specified in greater detail than is done by the authors. Second, it is argued that the class of local HVTs that still survives is wide and natural. In view of the far-reaching consequences these experiments may have for theoretical physics, it is essential to make this point.

Although the non-expert may believe that the experiments have definitely eliminated *all* local-realist theories – since the experiment [1] is claimed to be 'loophole-free' – upon closer inspection the authors of [1] do take precautions to specify their claim. Thus they state: "Strictly speaking, no Bell experiment can exclude all conceivable local-realist theories, because it is fundamentally impossible to prove when and where free random input bits and output values came into existence." And they conclude: "Our observation of a loophole-free Bell inequality violation thus rules out all local realist theories that accept that the number generators timely produce a free random bit and that the outputs are final once recorded in the electronics" (call this Claim 1). Regrettably, the authors only mention the locality and detection loopholes, and do not specify that their experiment is still vulnerable to the so-called 'freedom-of-choice' loophole, or an instance of the latter. This specification *is* made in Ref. [3]: "This [freedom-of-choice loophole] loophole can be closed only under specific assumptions about the origin of [the HVs] $\lambda$. Under the assumption that $\lambda$ is created with the particles to be measured, an experiment in which the settings are generated independently at the measurement stations and spacelike separated from the creation of the particles closes the loophole."

Now, it is certainly very natural to consider the $\lambda$ as pertaining only to the particle pair, more generally as created at the source with the particles, in the sense of [3]. And for such HVTs Claim 1 can indeed legitimately be inferred from the experiment. But it is not the only possibility that is logically allowed. The point of this Comment is that the class of HVTs still exploiting the freedom-of-choice loophole may be vast, and natural. For instance, the $\lambda$ could besides the particles also describe a background field or medium in which the particles move and that interacts with particles and analyzers [4, 8]. In this case the full dynamics of the $\lambda$ is crucial, as was recently investigated in detail in Ref. [4] (the background field may refer to vacuum fluctuations, the ether, a dark field,…). Specifically, let $\lambda \equiv (\lambda_0, \lambda_1, \lambda_2)$, where $\lambda_0$ are properties of the particle pair at emission in the sense of Ref. [3], $\lambda_1$ properties of the background field in the neighborhood of the left analyzer (with setting a), and $\lambda_2$ properties of the field close to the right analyzer (with setting b). Then it is clear that the conditional probability $P(\lambda|a,b) \equiv P(\lambda_0, \lambda_1, \lambda_2|a,b)$ is in general different from the unconditional $P(\lambda_0, \lambda_1, \lambda_2)$ simply because $\lambda_1$ can interact with analyzer 'a' and $\lambda_2$ with analyzer 'b'; therefore $\lambda$ can obviously be dependent on (a, b) even if the interactions can be entirely local. In such a 'background-based' HVT there *is* a probabilistic dependence between $\lambda$ and (a,b), but this in no way means that the settings (a,b) are conspiratorially determined by the $\lambda$. The input bits are free and random; there is just a stochastic correlation, as happens in countless probabilistic experiments [4]. That such background-based HVTs can reproduce the quantum correlation of the Bell experiment is shown in [4].

That the freedom-of-choice loophole is more subtle than meets the eye, has recently been argued by several researchers [8-11]. What is important for this Comment, independent of the details and even of the validity of the case study [4], is that there may be HVTs that do *not* assume "that λ is created with the particles to be measured". In our example $\lambda_1$ and $\lambda_2$ are field values of a background in the spacetime region of the measurement, not emission, events. In other words, it is well possible to devise "local realist theories that accept that the number generators timely produce a free random bit and that the outputs are final once recorded in the electronics", against Claim 1.

To give further physical grounds why in particular background-based HVTs are relevant in this discussion, it suffices to have a look at the spectacular experiments recently performed by teams led by Couder, Bush and Fort [5-7]. These researchers have shown that oil droplets can, under specific conditions, be made to walk over a vibrating fluid film, and mimic a wide range of quantum phenomena, including double-slit interference, orbit quantization and Zeeman splitting. The essential 'element of reality' that is responsible for such quantum-like behavior is the pilot or background wave that accompanies the droplets. The droplets hit the fluid film and create a surface wave on it, which guides their movement. In such fluid mechanical systems there is a complex dynamics between the (properties of the) droplets (the equivalent of $\lambda_0$ above), the background or pilot wave ($\lambda_1, \lambda_2$), and the detailed geometry of the fluid bath or any 'contextual' variables (a, b). Such systems exhibit massive correlations [7]. There are e.g. manifestly correlations of the type $P(\lambda_1, \lambda_2|a,b)$: the pilot wave is strongly dependent on, e.g., the geometry of the bath (yet (a,b) can be free or random variables). The probabilistic dependencies one has to assume in the background-based model for the Bell experiment [4], are of the same type as exist in the droplet systems [5-7].

Of course, the background-based theories we have in mind remind one of Louis de Broglie's pilot-wave theory. There is a whole community of physicists working on modern variants of this theory, attempting to derive quantum mechanics from a hidden level of reality (see a condensed review in [7]). One can also refer here to the cellular automaton theory of quantum mechanics of 't Hooft, which indeed features correlations as above [12].

In conclusion, the crucial experiment by Hensen et al. does eliminate, for the first time, an important class of local-realist models. But it does not eliminate a wide and natural class of local-realist theories. Similar remarks hold, mutatis mutandis, for Refs. [2-3]. In view of the importance of these experiments for physics – after all, the unification of quantum mechanics and relativity theory may be at stake – it is important to highlight this distinction.

*Acknowledgements. I thank Ronald Hanson for his kind willingness to discuss the experiment and its interpretation.*